\documentclass[journal]{IEEEtran}
\IEEEoverridecommandlockouts

\usepackage{tabularx}
\usepackage{pifont} % for symbols
\usepackage{ragged2e}

\usepackage{booktabs}

\usepackage{array}
\usepackage{etoolbox}
\usepackage[pdftex]{graphicx}
\graphicspath{{../pdf/}{../jpeg/}}
\DeclareGraphicsExtensions{.pdf,.jpeg,.png}

\usepackage[cmex10]{amsmath}
\usepackage{array}
\usepackage{mdwmath}
\usepackage{mdwtab}
\usepackage{eqparbox}
\usepackage{stfloats} % or \usepackage{dblfloatfix}

\usepackage{url}
\usepackage{hyperref} % For adding hyperlinks
\usepackage{orcidlink} % If you have the package for ORCID logo
\usepackage{cite}
\usepackage{mathrsfs}
\usepackage{amsmath,amssymb,amsfonts}
\usepackage{textcomp}
\usepackage{algorithm}
\usepackage{algpseudocode}
\usepackage{lipsum}
\usepackage{caption}
\usepackage{subcaption}
\usepackage{multirow}
\usepackage{nomencl}
\makenomenclature
\begin{document}
% \IEEEpubid{\makebox[\columnwidth]{979-8-3315-3544-5/25/\$31.00~\copyright2025 IEEE \hfill}
% \hspace{\columnsep}\makebox[\columnwidth]{ }}

\bstctlcite{IEEEexample:BSTcontrol}
 %   \title{ An Adaptive Detection and Mitigation Framework Against Stealthy Attacks in P2P Energy Trading}
 %  \title{ A Cyber-Resilient and Adaptive P2P Trading Framework Against Optimal Adversial Attacks}
 \title{An Adaptive Cyber-Resilient Framework for P2P Energy Trading under Optimal Stealthy Attacks}
  \author{
 Devangi,\,~\IEEEmembership{Graduate Student Member,~IEEE},   
    Yashasvi Bansal,\,~\IEEEmembership{Senior Member,~IEEE},
     Ankit Singhal,\,~\IEEEmembership{Senior Member,~IEEE}, \\ 
    Indian Institute of Technology Delhi, Hauz Khas, 110016\\
    Emails:- eez248399@ee.iitd.ac.in, yashasvi@ee.iitd.ac.in, sankit@ee.iitd.ac.in }

%\markboth{IEEE TRANSACTIONS ON MICROWAVE THEORY AND TECHNIQUES, VOL.~60, NO.~12, DECEMBER~2012
%}{Roberg \MakeLowercase{\textit{et al.}}: High-Efficiency Diode and Transistor Rectifiers}

% ====================================================================
\maketitle

% === ABSTRACT ====================================================================
% =================================================================================
\begin{abstract}

The increasing adoption of peer-to-peer (P2P) energy trading has enabled prosumers to actively participate in decentralized energy markets. However, the growing dependence of P2P energy markets on communication infrastructures and distributed coordination mechanisms introduces an expanded cyberattack surface, enabling sophisticated adversaries to manipulate market interactions while remaining concealed from conventional detection mechanisms. This paper first designs an optimal stealthy attack that maximizes economic disruption of the MicroGrids (MGs), while preserving the apparent convergence characteristics of the ADMM-based market-clearing process. To detect such attacks, a Mahalanobis–Boosting detector is developed using market-layer features to effectively capture inter-feature dependencies and distinguish compromised updates from normal market operation. An adaptive-rho based mitigation strategy is then introduced to limit the propagation of compromised information and restore robust market operation once an attack is flagged. The effectiveness of the proposed framework is demonstrated through comprehensive case studies, achieving an attack-detection accuracy of 98.49\% while the proposed mitigation strategy reduces the operating cost by approximately 18.43\% compared to the unmitigated attack case. The proposed approach provides a practical pathway toward trustworthy and cyber-secure distributed P2P energy trading systems.

\end{abstract}

% === KEYWORDS ====================================================================
% =================================================================================
\begin{IEEEkeywords}
Cyberattack, MicroGrids, P2P, Power Markets, Regression tree
\end{IEEEkeywords}

% \fi
\IEEEpeerreviewmaketitle

% ====================================================================
% ====================================================================
% ====================================================================
\renewcommand{\nomgroup}[1]{%
\item[\itshape
\ifstrequal{#1}{I}{Indices and Sets}{%
\ifstrequal{#1}{V}{Variables}{%
\ifstrequal{#1}{P}{Parameters}{%
\ifstrequal{#1}{F}{Functions}{%
}}}}%
]}

%%%%% Indices and sets 

\nomenclature[I]{$t$}{Time index}
\nomenclature[I]{$i$}{Number of MGs}
\nomenclature[I]{$k$}{Iteration index}
\nomenclature[I]{$l$}{Training sample index}
\nomenclature[I]{$o$}{K-means cluster centers}
\nomenclature[I]{$s$}{Regression tree}

%%%%%%Parameters

\nomenclature[P]{$a_i,b_i$}{Quadratic and linear  cost coefficient of MG $i$}
\nomenclature[P]{$\kappa_i$}{Battery cycling penalty parameter}
\nomenclature[P]{$\mathcal{D}_{i,t},\mathcal{R}_{i,t}$}{Local load demand and renewable generation}
\nomenclature[P]{$\mathcal{G}_i^{max}$}{Maximum dispatchable generation}
\nomenclature[P]{$B_i^{+,max}$}{Maximum charging power of battery}
\nomenclature[P]{$B_i^{-,max}$}{Maximum discharging power of battery}

\nomenclature[P]{$\mathcal{E}_i^{max},\mathcal{E}_i^{min}$}{Maximum/Minimum SOC}
\nomenclature[P]{$\eta_c,\eta_d$}{Charging and discharging efficiency}
\nomenclature[P]{$\Delta_i^{att}$}{Attack stealth bound}
 
%%%Variables 
\nomenclature[V]{$\mathcal{G}_{i,t}$}{Dispatchable generation of MG $i$ at time $t$}
\nomenclature[V]{$\mathcal{P}_{i,t}$}{Power traded in P2P market}
\nomenclature[V]{$B^+_{i,t},B^-_{i,t}$}{Charging and discharging power of battery}
\nomenclature[V]{$\mathcal{E}_{i,t}$}{State of charge of battery}
\nomenclature[V]{$z_{t},u_{t}$}{ADMM auxiliary consensus/scaled dual variable}
\nomenclature[V]{$\rho$}{ADMM penalty parameter}
\nomenclature[V]{$\delta^{att}_{i,t}$}{Attacker's injection at time $t$}
\nomenclature[V]{$\mathcal{C}_N, \mathcal{C}_A$}{K-means cluster centres }
\nomenclature[V]{$\mathcal{D}_N, \mathcal{D}_A$}{Minimum distance to the normal/attack center}
\nomenclature[V]{$\mathbf{\Sigma}_N, \mathbf{\Sigma}_A$}{Normal and attack covariance matrix}
\nomenclature[V]{$\psi_{\mathrm{attack}}$}{Probability of attack}

%%%%%Functions

%\renewcommand{\nomlabel}[1]{\makebox[1.8cm][l]{#1}}
%\printnomenclature
\renewcommand{\nomlabel}[1]{\makebox[1.8cm][l]{#1}}
\input{output1.nls}\textbf{}
% === I. INTRODUCTION =============================================================
% =================================================================================
\section{Introduction}
\label{section1}
The global transition toward a low-carbon and sustainable energy ecosystem has accelerated the deployment of distributed energy resources (DERs), transforming passive consumers into active prosumers and promoting decentralized energy systems in line with international decarbonization and clean energy initiatives \cite{UNSDG7,IPCC2023}. As renewable penetration continues to increase, conventional centralized market structures face growing challenges in efficiently coordinating geographically distributed resources. Consequently, peer-to-peer (P2P) energy trading has emerged as a promising decentralized market paradigm that enables direct energy exchange among prosumers, improving renewable utilization, economic efficiency, and the flexibility and resilience of modern distribution networks \cite{tushar2020peer}.

Despite their significant operational and economic advantage, P2P trading frameworks substantially expand the cyber attack surface due to their heavy reliance on DERs \cite{liu2024enhancing}, Internet of Things (IoT) devices, advanced metering infrastructure (AMI), communication networks, and cloud-based market coordination platforms, thereby increasing the number of potential cyber entry points available to malicious adversaries \cite{beg2023review}. Compromise of DER components, such as PV inverters \cite{li2022cybersecurity}, battery, EV charging stations, and smart meters, can manipulate local energy operations, while communication-layer attacks, including false data injection (FDI), replay, man-in-the-middle (MITM), and denial-of-service (DoS) attacks \cite{tatipatri2024comprehensive}, can disrupt information exchange and distort market-clearing decisions, thereby compromising market efficiency and grid stability.

%%% pAPERS DOING FDI IN POWER MARKETS:
Among various cyber threats, False Data Injection Attacks (FDIAs) have attracted significant research attention due to their ability to stealthily manipulate market outcomes and generate financial gains. Existing studies have investigated integrity attacks on real-time electricity pricing \cite{tan2015integrity}, ramp-induced data attacks \cite{choi2013ramp}, optimization-based FDI attacks under uncertainty \cite{farahani2024new}, parameter manipulation attacks \cite{xu2020power}, financially motivated attacks on real-time market operation \cite{liu2017financially}, worst-case data injection attacks \cite{mengis2017data}, market-level defense mechanisms against FDI and LMP manipulation \cite{zhang2020market}, and robust control strategies against cyber attacks in real-time electricity markets \cite{rouhani2025robust}. Collectively, these studies demonstrate that cyber attacks can significantly distort market clearing, electricity prices, and system operation, highlighting the need for effective market-level cyber attack detection mechanisms. Although several studies have investigated cyberattacks and defense mechanisms, limited attention has been devoted to optimization-based stealthy attacks in P2P energy trading.

Several attack detection mechanisms have been proposed to enhance the cyber-security of P2P energy trading. In \cite{liu2025byzantine}, a physics-informed spatio-temporal learning framework is proposed to identify Byzantine attacks. The detector is executed by each MG based on information exchanged directly with neighboring peers, in consistent with a decentralized peer-centric trading architecture as explained in \cite{xia2023reviewing}. In \cite{chang2022byzantine}, each agent mitigates Byzantine attacks by filtering out the most deviating variables (internal peer data) received from neighboring peers during the information exchange step of the distributed optimization process. Similarly, \cite{nguyen2021cooperative} employs a weighted-mean-subsequence-reduced (WMSR) resilient consensus algorithm that filters suspicious consensus states received from neighboring prosumers to maintain resilient market consensus. While \cite{mohammadi2021detecting} identifies anomalies by analyzing household energy consumption patterns using clustering and correlation-based similarity measures. However, these approaches typically rely on internal MG or household-level information, such as neighboring states and local measurements, which may not be readily accessible in distributed P2P energy markets. In such settings, MGs preserve operational privacy and coordinate primarily through market-layer information rather than exchanging detailed internal states, thereby limiting the direct applicability of these detection mechanisms.

%In \cite{mohamed2021secured}, a modified Intelligent Priority Selection-based Reinforcement Learning (IPS-RL) algorithm detects false data injection attacks on measurement data.
\vspace{-0.2cm}
\subsection{Challenges and Contributions}
Based on the above literature review, some key research challenges are: (i) cyber attacks should be capable of simultaneously maximizing economic disruption while remaining difficult to detect (stealthy), (ii) accurately detecting stealthy cyber attacks at the market layer of distributed P2P energy trading (using only the features available at the market layer), (iii) effectively incorporating attack-probability information into the market-clearing process to adaptively mitigate the influence of suspicious information. To address the aforementioned challenges, the main contributions of this paper are  as follows:
\begin{enumerate}
    \item \textit{Attack:} In contrast to \cite{tan2015integrity}--\cite{rouhani2025robust}, a stealth-aware optimal FDI attack is modeled for P2P energy trading that maximizes market disruption while satisfying stealthy characteristics.
    \item \textit{Detection:} In contrast to existing detection approaches \cite{liu2025byzantine}--\cite{mohammadi2021detecting}, a Mahalanobis--Boosting detection framework is developed to detect stealthy attacks using only the variables available at the market layer, where Mahalanobis distances capture the inter-feature dependencies and are subsequently processed by a Gradient Boosting classifier to learn their nonlinear relationships and distinguish stealthy attacks from normal operating conditions.
    \item \textit{Mitigation:} Instead of hard rejection of suspicious communicated variables as in \cite{chang2022byzantine}--\cite{nguyen2021cooperative}, a cyber-aware adaptive ADMM mechanism is proposed in which the detector dynamically adjusts the ADMM penalty parameter according to the estimated attack probability.
\end{enumerate}
\subsection{Proposed System Framework}
\begin{figure}[h!]
\vspace{-0.4cm}
    \centering
    \includegraphics[width=0.85\linewidth]{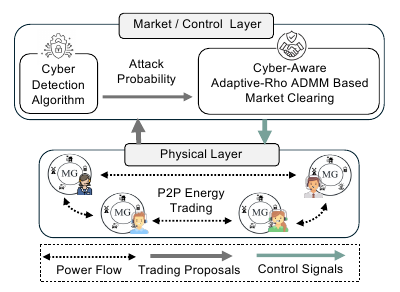}
    \caption{Proposed System Framework}
    \label{fig:frame}
    \vspace{-0.2cm}
\end{figure}

The proposed framework comprises two interconnected layers, namely the physical layer and the market/control layer, as illustrated in Fig.~\ref{fig:frame}.
\begin{enumerate}
    \item \textit{Physical Layer:} MGs equipped with DERs participate in P2P energy trading by exchanging energy locally. During each market-clearing iteration, the MGs transmit the proposed P2P trades to the market layer and receive the updated control variables required for the subsequent ADMM iteration.

    \item \textit{Market/Control Layer:} The market/control layer comprises two functional modules: (i) a cyber detection module that continuously monitors the proposed P2P trading information exchanged during market coordination and estimates the probability of stealthy cyber attacks, and (ii) an ADMM-based market-clearing module that incorporates the estimated attack probability through an adaptive-rho mitigation strategy to regulate the coordination process and compute the updated control variables and market-clearing decisions.
\end{enumerate}
Overall, the proposed framework tightly integrates cyber attack detection with distributed market clearing, thereby enabling secure, adaptive, and cyber-resilient P2P energy trading.
The remainder of this paper is organized as follows. Section \ref{Section2} presents the system model and attack formulation. Sections \ref{section3} and \ref{section4} describe the proposed detection framework and attack-aware adaptive ADMM algorithm, respectively. Section \ref{section5} presents the test system and dataset, Section \ref{section6} discusses the results, and Section \ref{section7} concludes the paper. The 
\section{Mathematical Modeling} 
\label{Section2}
This section describes the mathematical formulation of the MG optimization, the distributed market-clearing mechanism, and the construction of the trading graph used to represent P2P energy transactions and the attacker's optimization.
\subsection{Mathematical Modeling of MicroGrids (MGs)}
Consider a P2P market consisting of $M$ interconnected MGs \cite{ali2022economic}, consisting of load, solar, DGs and battery. For each MG $i \in \{1,...,M\}$ at time $t$, the local operating cost ($\mathcal{J}_{i,t}$), which the MG seeks to minimize, is defined as
\begin{equation}
\mathcal{J}_{i,t}=a_i \mathcal{G}_{i,t}^2 + b_i \mathcal{G}_{i,t} + \kappa_{i} \big( (B^+_{i,t})^2 + (B^-_{i,t})^2 \big),
\label{eq:local_objective}
\end{equation}
where \(a_i \mathcal{G}_{i,t}^2 + b_i \mathcal{G}_{i,t}\) represents the DG generation cost, and \(\kappa_i \big( (B^+_{i,t})^2 + (B^-_{i,t})^2 \big)\) denotes the degradation cost associated with charging and discharging of battery. The above objective is subject to the following operational constraints.
\begin{equation}
\mathcal{G}_{i,t} = \mathcal{D}_{i,t} + \mathcal{P}_{i,t} - \mathcal{R}_{i,t} + B^+_{i,t} - B^-_{i,t}
\label{eq:power_balance}
\end{equation}
\begin{equation}
\sum_{i=1}^{M} \mathcal{P}_{i,t} = 0
\label{eq:market_clearing}
\end{equation}
\begin{equation}
0 \le \mathcal{G}_{i,t} \le \mathcal{G}_i^{\max}, 0 \le B^+_{i,t} \le B_i^{+,\max}, 
0 \le B^-_{i,t} \le B_i^{-,\max} 
\label{eq:g_limits}
\end{equation}
\begin{equation}
\mathcal{E}_{i,t+1}
=
\mathcal{E}_{i,t}
+
\eta_c B^+_{i,t}
-
\frac{1}{\eta_d} B^-_{i,t}; \quad \mathcal{E}_i^{\min}
\le 
\mathcal{E}_{i,t}
\le
\mathcal{E}_i^{\max}
\label{eq:soc_dynamics}
\end{equation}
where, Eq. \ref{eq:power_balance} is the power balance constraint, Eq. \ref{eq:market_clearing} is the constraint ensuring market clearing, Eq. \ref{eq:g_limits} are the operational constraints, Eq. \ref{eq:soc_dynamics} is the SOC dynamics of the battery. The overall objective function of MG can be written as.
\begin{equation}
\min_{\{\mathcal{G}_{i,t},\mathcal{P}_{i,t},B^+_{i,t},B^-_{i,t}\}}
\sum_{i=1}^{M} \mathcal{J}_{i,t}
\label{eq:compact_centralized}
\end{equation}
\begin{equation*}
    s.t. (\ref{eq:power_balance}) - (\ref{eq:soc_dynamics})
\end{equation*}
The optimization problem in \eqref{eq:compact_centralized} is reformulated using the ADMM \cite{nguyen2020optimal}, \cite{jin2025peer}, which enables distributed optimization while enforcing the global market-clearing constraint through coordination among MGs.
\vspace{-0.3cm}
\subsection{ADMM Reformulation}
\label{sectionadmm}
To enable distributed optimization, an auxiliary consensus variable 
($z \in \mathbb{R}^{M}$) is introduced, and the market clearing constraint \eqref{eq:market_clearing} is reformulated as
\begin{equation}
\mathcal{P}_{t} - z_{t} = 0  \hspace{0.2cm} \text{and} \hspace{0.2cm} \sum_{i=1}^{M} z_{i,t} = 0
\label{eq:admm_consensus}
\end{equation}
Introducing a dual variable ($\lambda$) and  a penalty parameter ($\rho > 0$) for the constraint \eqref{eq:admm_consensus}, defining the scaled dual variable as $u = \frac{\lambda}{\rho} $ and substituting $\lambda = \rho u$, the augmented Lagrangian $\mathcal{L}_{\rho}(\mathcal{P}_{t},B^+_{t},B^-_{t},z_t,u_t)$ is given by:
\begin{equation}
\sum_{i=1}^{M} \mathcal{J}_{i,t}
+
\frac{\rho}{2}
\|\mathcal{P}_t - z_t + u_t\|_2^2.
\label{eq:scaled_augmented_lagrangian}
\end{equation}
At iteration $k$, the updates proceed as follows.
\subsubsection{\textit{Primal Update}}
For each MG $i$, $\mathcal{P}_{i,t}^{k+1},\, (B^+_{i,t})^{k+1},\, (B^-_{i,t})^{k+1}$ is equal to: 
\begin{equation}
\begin{aligned}
\arg\min_{\mathcal{P}_{i,t},\,B^+_{i,t},\,B^-_{i,t}} 
&
\mathcal{J}_{i,t}
+
\frac{\rho}{2}
\left( \mathcal{P}_{i,t} - z_{i,t}^{k} + u_{i,t}^{k} \right)^2
\end{aligned}
\label{eq:primal_update}
\end{equation}
Let $\zeta_i \in \{+1,-1\}$ indicate charging/discharging of battery. Then
\begin{equation}
B_{i,t}^{k+1}
=
\frac{1+\zeta_i}{2}\, (B^+_{i,t})^{k+1}
+
\frac{1-\zeta_i}{2}\, (B^-_{i,t})^{k+1}.
\end{equation}
Calculating $\delta\mathcal{L}/\delta \mathcal{P}_{i,t}, \delta\mathcal{L}/\delta B^+_{i,t}$ and $ \delta\mathcal{L}/\delta B^-_{i,t}$ and equating to zero yields the following closed-form update.
\begin{equation}
\begin{aligned}
\begin{bmatrix}
\mathcal{P}_{i,t}^{k+1} \\
B_{i,t}^{k+1}
\end{bmatrix}
&=
\Pi_{\Omega_{i,t}}
\Bigg(
A^{-1} \times
\begin{bmatrix}
-\theta_{i,t} + \rho\big(z_{i,t}^k - u_{i,t}^k\big) \\
-\sigma_i \theta_{i,t}
\end{bmatrix}
\Bigg).
\end{aligned}
\end{equation}
\begin{equation*}
\text{where}
\quad
A =
\begin{bmatrix}
2a_i + \rho & 2a_i \sigma_i \\
2a_i \sigma_i & 2a_i + 2\kappa_i
\end{bmatrix}.
\end{equation*}
and, $\theta_{i,t} = 2a_i \big(\mathcal{D}_{i,t} - \mathcal{R}_{i,t}\big) + b_i$ , $\Pi_{\Omega_i(t)}(\cdot)$ denotes the euclidean projection onto the feasible set $\Omega_i(t)$ (\eqref{eq:g_limits}).
\subsubsection{\textit{Consensus ($z$) Update}} The z-update is used to enforce market clearing (global balance) in a distributed way.
\begin{equation}
z_{t}^{k+1}
=
\arg\min_{z_{t}}
\frac{\rho}{2}
\left\| \mathcal{P}_{t}^{k+1} - z_{t} + u_{t}^{k} \right\|_2^2
\quad
\text{s.t. }
\sum_{i=1}^{M} z_{i,t} = 0
\label{eq:z_update}
\end{equation}
Solving \eqref{eq:z_update} using Lagrange multipliers yields.
\begin{equation}
z_{i,t}^{k+1}
=
\mathcal{P}_{i,t}^{k+1}
+
u_{i,t}^{k}
-
\frac{1}{M}
\sum_{j=1}^{M}
\left( \mathcal{P}_{j,t}^{k+1} + u_{j,t}^{k} \right)
\end{equation}
\subsubsection{\textit{Dual Update}}The dual update is written as:
\begin{equation}
u_{i,t}^{k+1}
=
u_{i,t}^{k}
+
\mathcal{P}_{i,t}^{k+1}
-
z_{i,t}^{k+1}.
\label{eq:dual_update}
\end{equation}
\subsubsection{\textit{Convergence Criterion}}
The primal residual ($r_t$) and the dual residual ($s_t$) is defined as
\begin{equation}
r_t^{k} = \mathcal{P}_t^{k} - z_t^{k}, \quad s_t^k = \rho \left( z_t^k - z_t^{k-1} \right),
\label{eq:primal_residual}
\end{equation}
and convergence is achieved when $\|r_t^k\|_2 \leq \varepsilon^{\text{pri}}, \quad \|s_t^k\|_2 \leq \varepsilon^{\text{dual}},$
% \begin{equation}
% \|r_t^k\|_2 \leq \varepsilon^{\text{pri}}, \quad \|s_t^k\|_2 \leq \varepsilon^{\text{dual}},
% \label{tolerance}
% \end{equation}
where $\varepsilon^{\text{pri}}, \varepsilon^{\text{dual}}$ is a prescribed tolerance.

\subsection{Building a Trading Graph}
\label{tradingraph}
Following convergence, the cleared net positions are mapped into quantity-preserving bilateral transactions $\mathcal{P}_{ij,t}$ through 
the market's trade-matching mechanism, such that $\mathcal{P}_{i,t}
=
\sum_{j\neq i}\mathcal{P}_{ij,t};
\mathcal{P}_{ij,t}=-\mathcal{P}_{ji,t}$.
At each time interval $t$, the P2P trading market is modeled as a weighted directed graph $G_t=(V,E,W_t)$, 
constructed using the bilateral P2P trade variables ($\mathcal{P}_{i,j}$) (communicated by the MGs to the control layer during the market-clearing process). The graph components are defined as
\begin{equation}
V=\{MG_1,MG_2,\ldots,MG_M\},
\end{equation}
\begin{equation}
E=\{(i,j)\mid i,j\in V,\; i\neq j\}, \quad W_t=\left[\mathcal{P}_{ij,t}\right]_{M\times M},
\end{equation}
Here, $V$ denotes the set of participating MGs, $E$ represents the set of P2P trading links, $W_t$ denotes the weighted adjacency matrix of the P2P trading graph at time interval $t$, and $\mathcal{P}_{ij,t}$ denotes the P2P power traded from MG$_i$ to MG$_j$. The cumulative bilateral transaction matrix is obtained as $F=\sum_{t=1}^{T}W_t=[F_{ij}]_{M\times M} =\sum_{t=1}^{T}\mathcal{P}_{ij,t}$
\subsection{Attacker's Optimization}  
\label{SECTIONIV}  
An adversarial agent is introduced to assess the vulnerability of the P2P market by injecting false data into the P2P trading information communicated between the compromised MGs and the market-clearing layer (attacker's entry point). The attacker seeks to maximize disruption while maintaining stealth. The objective function of the attacker ($\mathcal{W}_{i,t}$) can be written as: 
% \begin{equation}
% \begin{aligned}
% \mathcal{W}_{t}= \quad
% & \lambda_1 \sum_{i\in \mathcal{G}}\left( \mathcal{J}_{i,t}^{att}- \mathcal{J}_{i,t}\right)
% - \lambda_2 \sum_{i \in \mathcal{G}} \left| \delta_{i,t}^{att} - \delta_{i,t-1}^{att} \right|
% \end{aligned}
% \end{equation}
\begin{equation}
\begin{aligned}
\mathcal{W}_{t}= \quad
& \sum_{i\in \mathcal{G}}\left( \mathcal{J}_{i,t}^{att}- \mathcal{J}_{i,t}\right)
\end{aligned}
\end{equation}
where $\mathcal{J}_{i,t}$ and $\mathcal{J}_{i,t}^{att}$ is the cost of MG $i$ under normal cases and attacked case (Since the adversary is assumed to have
prior access to the EMS information of the compromised MGs), and $\mathcal{G}$ are the MGs with attacked P2P transactions.
\begin{equation}
\tilde{\mathcal{P}}_{i,t} = \mathcal{P}_{i,t} + \delta^{att}_{i,t}, \quad \delta_{i,t}^{att} = 0, \quad \forall i \notin \mathcal{G}
\label{att}
\end{equation}
\begin{equation}
\sum_{i=1}^{M} \delta_{i,t}^{att} = 0; 
\label{ref:attmc}
\end{equation}
% \begin{equation}
%     -\rho_i^{att}\leq (\delta_{i,t}^{att}- \delta_{i,t-1}^{att}) \leq \rho_i^{att}; \quad -\bar{\delta}_{i} \leq \delta_{i,t}^{att} \leq \bar{\delta}_{i}
% \label{attackbound}
% \end{equation}
\begin{equation}
    \left| \delta_{i,t}^{att} - \delta_{i,t-1}^{att} \right| 
\leq \Delta_i^{att}; \quad \sum_{i\in\mathcal G}|\delta_{i,t}^{att}|\leq B^{att}
\qquad i \in \mathcal{G}.
\label{attackbound}
\end{equation}
\begin{equation}
\mathcal{P}_i^{\min} \leq (\mathcal{P}_{i,t} + \delta_{i,t}^{att} )\leq \mathcal{P}_i^{\max}, 
\label{range}
\end{equation}
Eq. \ref{att} represents the modified net injection under attack.
After the attack, the market-clearing residual  $\sum_{i=1}^{N} (\mathcal{P}_{i,t} + \delta_{i,t}^{att})= 0$ becomes identical to the market-clearing condition in Eq.~\eqref{eq:market_clearing}. Consequently, equating the two expressions yields Eq. \ref{ref:attmc}. The attack stealth and attack budget is constrained by Eq. \eqref{attackbound}, where \( \delta_{i,t}^{att} \) and \( \delta_{i,t-1}^{att} \) denotes the injected perturbation at time $t$ and $t-1$ respectively, while the bound on the attacked trade is shown in Eq. \ref{range}. 
\begin{figure*}[t]
    \centering
    \includegraphics[width=0.88\linewidth]{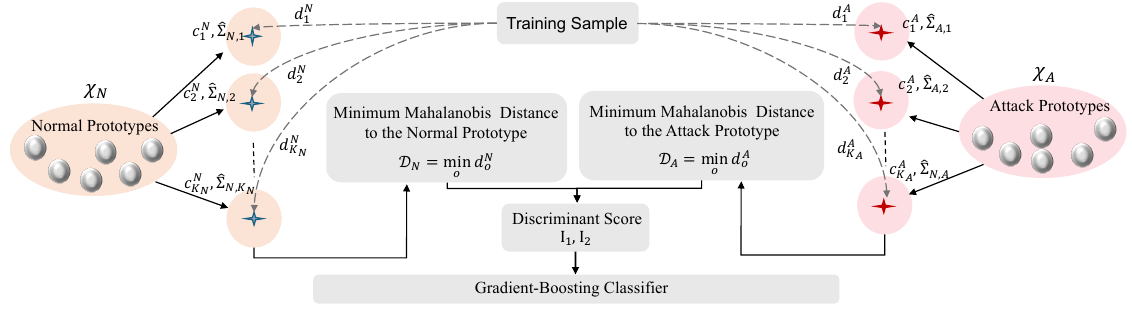}
    \caption{Mahalanobis feature extraction framework for cyber attack detection.}
    \label{fig:distnace}
\end{figure*}
The overall objective function of the attacker can be written as:
\begin{equation}
\max_{\delta_{i,t}^{att}} \quad 
 \mathcal{W}_{i,t}
\end{equation}
\begin{equation*}
    s.t. (\ref{att}) - (\ref{range})
\end{equation*}
\section{Proposed Mahalanobis--Boosting Detection}
\label{section3}
Consider a P2P energy trading market, during each market-clearing interval, the market layer observes only the P2P power generated by each MG, while private DER-level measurements remain inaccessible. Therefore, the feature vector, available to the control layer is: 
\begin{equation}
\mathbf{x}_{l}
=
\left[
r_{l},
s_{l},
u_{l},
\phi(G_{l})
\right],
\label{inpdetect}
\end{equation}   
where $l \in \{1,...,L\}$ denotes the training sample index, with $L$ representing the total number of training market-clearing samples, $r_l$ and $s_l$ are the primal and dual residual, $u_l$ is the scaled dual variable [Section \ref{sectionadmm}] and $\phi(G_{l})$ are the graph features for each sample $l$ derived from the trading graph [Subsection \ref{tradingraph}]. The standardized feature vector is $\tilde{\mathbf{x}}_l$.   

Stealthy attacks preserves the residual behavior while subtly distorting the inter-feature dependencies that characterize normal market operation. To capture these dependencies, class-specific Mahalanobis distance features are constructed, where the prototype centers characterize representative normal and attack operating patterns, while the covariance information captures the relationships among market-layer features.

%With respect to the changing graph features (due to attack), the residuals and dual variables remain largely unchanged, due to the preserved ADMM convergence of stealthy attack.
%To address this, Mahalanobis-distance-based statistical features are extracted by accounting for the covariance structure of the communicated variables, thereby highlighting subtle multivariate deviations and improving class separability.

\vspace{-0.4cm}
\subsection{K-means clustering}
The standardized normal training samples $\mathcal{\chi}_N$, where \(N\) is the number of normal samples, are first collected. The normal dataset is partitioned into $K_N$ clusters using the K-means algorithm \cite{steinley2006k}, which is performed only on the training dataset, by solving
\begin{equation}
\min_{\{\mathbf{c}_o^N\}}
\sum_{\tilde{\mathbf{x}}_l\in\mathcal{\chi}_N}
\min_{o}
\left\|
\tilde{\mathbf{x}}_l-\mathbf{c}_o^N
\right\|_2^2,
\quad
o\in\{1,\ldots,K_N\}.
\end{equation}
yielding the centers of the clusters as follows: $\mathcal{C}_N=
\left\{
\mathbf{c}_1^N,
\mathbf{c}_2^N,
\ldots,
\mathbf{c}_{K_N}^N
\right\}$
Similarly, the attack dataset $\mathcal{\chi}_A$ where \(A\) is the number of attack samples, partitioned into $K_A$ clusters using the K-means algorithm, yielding $    \mathcal{C}_A=
\left\{
\mathbf{c}_1^A,
\mathbf{c}_2^A,
\ldots,
\mathbf{c}_{K_A}^A
\right\}$

\subsection{Covariance Regularization}
To obtain a robust estimate of the covariance matrix, the Ledoit--Wolf \cite{ledoit2004well} shrinkage estimator is employed. The sample covariance matrix may become ill-conditioned due to feature correlations, resulting in unstable mahalanobis distance computations. The Ledoit–Wolf regularized covariance matrices $\hat{\mathbf{\Sigma}}_{c,o}$ is calculated as:
\begin{equation}
\hat{\mathbf{\Sigma}}_{c,o}
=
(1-\alpha_{c,o})\mathbf{S}_{c,o}
+
\alpha_{c,o}\mathbf{T}_{c,o},\quad c \in \{N, A\}.
\end{equation}
where $\mathbf{S}$ is the sample covariance matrix, $\mathbf{T}$ is the shrinkage target, and $\alpha\in[0,1]$ is the shrinkage coefficient. The covariance matrices for the normal and attack operating regions are denoted by $
\Sigma_N =
\left\{
\widehat{\Sigma}_{N,1}, 
\widehat{\Sigma}_{N,2}, 
\ldots,
\widehat{\Sigma}_{N,K_N}
\right\}$
 \text{and} 
$ \Sigma_A =
\left\{
\widehat{\Sigma}_{A,1},
\widehat{\Sigma}_{A,2},
\ldots,
\widehat{\Sigma}_{A,K_A}
\right\}
$ respectively.
\subsection{Mahalanobis Distance}
Since the communicated market variables are highly correlated, the Mahalanobis distance is employed to account for their covariance structure. For each sample $\tilde{\mathbf{x}}_l$, the mahalanobis distance \cite{de2000mahalanobis} to every normal prototype is computed. 
\begin{equation}
d_o^N=
(\tilde{\mathbf{x}}_l-\mathbf{c}_o^N)^T
\mathbf{\Sigma}_{N,o}^{-1}
(\tilde{\mathbf{x}}_l-\mathbf{c}_o^N).
\end{equation} 
\begin{equation}
\mathcal{D}_N=\min_o d_o^N,
\quad
o\in\{1,\ldots,K_N\}.
\end{equation}
Similarly, the distance to the attack prototypes $\mathcal{D}_A$ is:
\begin{equation}
\min_o
(\tilde{\mathbf{x}}_l-\mathbf{c}_o^A)^T
\mathbf{\Sigma}_{A,o}^{-1}
(\tilde{\mathbf{x}}_l-\mathbf{c}_o^A),
\quad
o\in\{1,\ldots,K_A\}.
\end{equation}
The following indicators are then constructed:
\begin{equation}
\mathbf{I_1}=\mathcal{D}_N-\mathcal{D}_A, \mathbf{I_2}=\frac{\mathcal{D}_N}{\mathcal{D}_A+\varepsilon}, 
\label{SGR}
\end{equation}
where, $\mathbf{I_1}$ is the distance contrast and $\mathbf{I_2}$ is the distance ratio between the normal and attack class geometries.
\begin{figure}[b]
\vspace{-0.2cm}
    \centering
    \includegraphics[width=\linewidth]{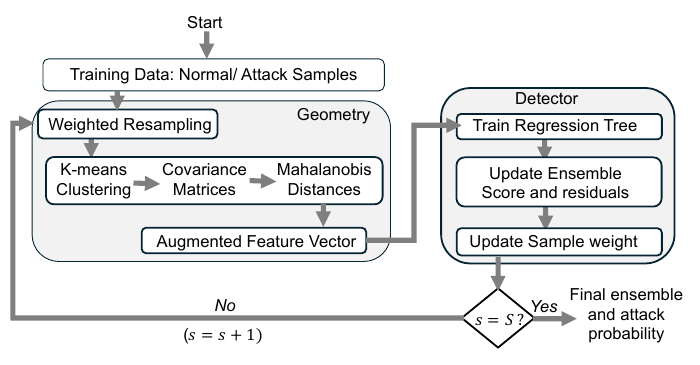}
    \caption{Training architecture of the proposed Mahalanobis–Boosting attack detection framework.}
    \label{fig:training}
    \vspace{-0.5cm}
\end{figure}
\vspace{-0.6cm}
\subsection{Gradient Boosting Classification}
The augmented feature vector $\mathbf{z}$ is fed inside the gradient boost classifier \cite{friedman2001greedy} (consisting of $S$ regression trees) to learn the non-linear relationship among these features. $\mathbf{z}$ is formed by concatenating the original feature vector in Eq.\ref{inpdetect}, with the features defined in Eq.\ref{SGR}. The ensemble prediction after $S$ boosting stages is given by:
\begin{equation}
F(\mathbf{z})= F_0
+\sum_{s=1}^{S}
\eta h_s(\mathbf{z}),
\end{equation}
where, $h_s(\cdot)$ denotes the $s^{\mathrm{th}}$ regression tree, $S$ is the total number of boosting stages, and $\eta$ is the learning rate. Each subsequent tree is trained to fit the negative gradient of the binary classification loss $L$, that is the pseudo-residual $r_l^{(s)}$,
\begin{equation}
r_l^{(s)}=-
\frac{\partial L_{loss}}
{\partial F_l^{(s-1)}(\mathbf{z})},
\label{res}
\end{equation}
where $L_{loss}$ denotes the binary cross-entropy (log-loss) used to train the gradient boosting classifier, $F_l^{(s-1)}$ represents the ensemble prediction of the $l^{\mathrm{th}}$ training sample obtained after the $(s-1)^{\mathrm{th}}$ boosting stage. After completion of all boosting stages, the ensemble output is converted into the attack probability through the sigmoid activation function,
\begin{equation}
\psi_{\mathrm{attack}}
=
\sigma(F(\mathbf{z})),
\end{equation}

\begin{algorithm}[!b]
\caption{Training of the Proposed Mahalanobis-Boosting Detector}
\label{alg:EGLW}
\begin{algorithmic}[1]
\Require
Training samples
$\{(\mathbf{x}_l,y_l)\}_{l=1}^{L}$, where $y_l\in\{0,1\}$ is the class label, number of prototypes $K_N,K_A$, boosting stages $S$, learning rate $\eta$
\State Standardize $\mathbf{x}_l\rightarrow\widetilde{\mathbf{x}}_l$
\State Initialize $F_l^0= \log\left(\frac{\pi}{1-\pi}\right),$
\For{$s=1,\ldots,S$}
\State Compute the current ensemble attack probabilities. 
\[ \psi_{\mathrm{attack},l}^{(s)}= \sigma\!\left(F_l^{(s-1)}\right),\quad l=1,\ldots,L\]
    \State Compute pseudo-residual as given in Eq. \ref{res}.
    \State Compute the hardness scores and normalized weights:
    \[
    q_l^{(s)}
    =
    \left|r_l^{(s)}\right|+\delta,
    \qquad
    w_l^{(s)}
    =
    \frac{q_l^{(s)}+\varepsilon}
    {\sum_{j=1}^{L}\left(q_j^{(s)}+\varepsilon\right)}
    \]
    \State Generate hardness-weighted datasets
     $\widetilde{\chi}_N^{(s)}, \widetilde{\chi}_A^{(s)}$
    \State Update centers and  covariance matrices
    \[ \mathcal C_N^{(s)},\mathcal C_A^{(s)},\Sigma_N^{(s)}, \Sigma_A^{(s)}\]
    \State Compute the Mahalanobis distance 
     $\mathcal{D}_{N,l}^{(s)},\mathcal{D}_{A,l}^{(s)}$
        \State Construct the augmented feature vectors:
    \[
    \mathbf z_l^{(s)}
    =
    \Big[
    \widetilde{\mathbf x}_l,\,
    (\mathbf{I_1})_l^{(s)},\,
    (\mathbf{I_2})_l^{(s)},\,
    (\mathbf{I_3})_l^{(s)}\
    \Big]
    \]
    \State \State Train regression tree $h_s(\mathbf z)$ using
$
\left\{
\left(\mathbf z_l^{(s)},r_l^{(s)}\right)
\right\}_{l=1}^{L}
$
    \State Update ensemble\[F_l^{(s)}=F_l^{(s-1)}+ \eta h_s\!\left(\mathbf z_l^{(s)}\right),\quad l=1,\ldots,L\]
    \State Store all stage-specific prototypes, covariance matrices, and regression trees.
\EndFor
\State For a test sample, evaluate the stored ensemble to obtain the attack probability.
\State Classify the sample as attack if $\psi_{\rm attack}\ge\tau$;
otherwise classify it as normal.
\end{algorithmic}
\end{algorithm}
where $\sigma(\cdot)$ is the sigmoid function. Finally, the attack probability is compared against an optimal decision threshold selected by maximizing the F1-score, thereby achieving the best trade-off between attack detection and false alarms.
\subsection{Overall Algorithm}
Fig. \ref{fig:training} and Algorithm \ref{alg:EGLW} summarize the training procedure of the proposed  Mahalanobis–Boosting detector. Here, $\pi$ denotes the prior attack probability estimated from the training data and used to initialize the ensemble prediction, $\epsilon$ is a small positive constant introduced for numerical stability during weight normalization. At each boosting stage, the magnitude of the pseudo-residual is used to quantify the hardness of each training sample. Samples that are poorly represented by the current ensemble are assigned larger feedback weights and are therefore sampled more frequently during the geometry reconstruction process. Consequently, the class-specific prototypes and covariance matrices are iteratively updated to better capture the difficult regions of the feature space, resulting in an Mahalanobis distance representation. These geometric features are then concatenated with the original feature vector to construct an augmented feature representation, on which a regression tree is trained to update the ensemble prediction. Refining the prototype geometry at each boosting stage improves normal–attack class separability and subsequent detection performance.
\vspace{-0.3cm}
\section{Attack Aware Adaptive ADMM}
\label{section4}
To mitigate the propagation of compromised communication updates in ADMM, a detector-guided attack-aware adaptive-rho framework is proposed. Unlike conventional adaptive ADMM schemes \cite{boyd2011distributed,gao2018information}, which update rho using the primal and dual residuals, the proposed approach adapts rho according to the estimated attack probability while preserving the ADMM framework. At iteration $(k)$, the detector estimates the attack probability $(\psi_{\mathrm{attack}}^{k})$ from the received trading vector. The normalized detector confidence is defined as
\begin{equation}
q^{k} 
=
\frac{\psi_{\mathrm{attack}}^{k}-\tau_{\mathrm{L}}}
{\tau_{\mathrm{H}}-\tau_{\mathrm{L}}},
\label{eq:normalized_confidence}
\end{equation}
where $q^k \in [0,1)$ increases with the estimated attack probability and controls the degree of penalty adaptation, $\tau_{\mathrm{L}}=\tau-0.1$ and $\tau_{\mathrm{H}}=\tau+0.1$ denote the lower and upper security thresholds, respectively, defined as a $\pm0.1$ margin around the decision threshold $\tau$. If the communicated update is  normal ($P_{\mathrm{attack}}^{k}<\tau_{\mathrm{L}}$) then accepted, if suspicious ($\tau_{\mathrm{L}}\leq P_{\mathrm{attack}}^{k}<\tau_{\mathrm{H}}$) then adaptively processed, and if attacked ($P_{\mathrm{attack}}^{k}\geq\tau_{\mathrm{H}}$) then replaced by the most recent trusted trading vector, thereby mitigating compromised communications while preserving the standard ADMM framework.
\begin{algorithm}[b]
\caption{ADMM-Based Distributed P2P Market Clearing}
\label{alg:admm}
\begin{algorithmic}[1]
\State Initialization: Detector-estimated attack probability $P_{\mathrm{attack}}^{k}$ (obtained from Algorithm~\ref{alg:EGLW}), $\varepsilon^{\mathrm{pri}}$ and $\varepsilon^{\mathrm{dual}}$.
\For{$t=1$ to $24$}
\While{$\|r^{k}\|_{2}>\varepsilon^{\mathrm{pri}}$,
$\|s^{k}\|_{2}>\varepsilon^{\mathrm{dual}}$
\textbf{and} $k<K_{\max}$}
     \State Update local decision variables using (\ref{eq:primal_update}).
     \State Classify $\psi_{\mathrm{attack}}^{k}$ as normal, suspicious, or attacked
     \State Update $\rho$ according to adaptive-rho (Eq. \ref{eq:rho_update}).
\State Update consensus variable using (\ref{eq:z_update}).
\State Update scaled dual variable using (\ref{eq:dual_update}).
\EndWhile
%\State Store optimal objective value, P2P trades, residuals, and dual variables.
\EndFor
\State Output: 24-hour market-clearing results.
\end{algorithmic}
\end{algorithm}
Then, the target ADMM penalty parameter is determined using following equation.
\begin{equation}
\rho_{\mathrm{tar}}^{k}=
\begin{cases}
\rho_{0},
& \psi_{\mathrm{attack}}^{k}<\tau_{\mathrm{L}},\\[2mm]
\rho_{0}\left(1+\alpha_{\rho}q^{k}\right),
& \tau_{\mathrm{L}}\leq \psi_{\mathrm{attack}}^{k}<\tau_{\mathrm{H}},\\[2mm]
\rho_{\max},
& \psi_{\mathrm{attack}}^{k}\geq\tau_{\mathrm{H}}.
\end{cases}
\label{eq:rho_target}
\end{equation}
where, $\alpha_{\rho}=\dfrac{\rho_{\max}-\rho_{0}}{\rho_{0}}$, finally, the ADMM penalty parameter is updated according to
\begin{equation}
\rho^{k+1}
=
\lambda_{\rho}\rho^{k}
+
\left(1-\lambda_{\rho}\right)\rho_{\mathrm{tar}}^{k},
\label{eq:rho_update}
\end{equation}
where, the smoothing factor $\lambda_{\rho} \in [0,1]$ specifies the memory of the penalty update, determining the contribution of the previous penalty parameter relative to the target penalty. A larger value retains more memory, while a smaller increase the responsiveness to attack. The overall ADMM-based distributed market-clearing procedure is summarized in Algorithm \ref{alg:admm}. Consequently, the proposed detector-guided penalty adaptation enables the optimization process to dynamically respond to the estimated attack confidence. The proposed adaptive ADMM retains the convergence properties of the underlying ADMM under bounded penalty adaptation and admissible trusted-iterate perturbations, following established variable-step and inexact proximal-point convergence results \cite{rockafellar1976monotone},\cite{solodov2000inexact}.
\section{Test System and Dataset}
\label{section5}
The distributed MG optimization was implemented using an ADMM-based algorithm in NumPy, while the attack perturbation was determined at each ADMM iteration using the local constrained  Sequential Least Squares Programming (SLSQP) solver. Simulations were performed in the Anaconda environment using Spyder IDE (Python 3.12.2) on a MacBook Pro (Apple M4, 16 GB RAM, 10-core GPU) running macOS Sequoia 15.4.1.
 
\subsection{MicroGrid and Attacker's Specification}
Consider a P2P market consisting of $M=8$ interconnected MGs, consisting of load, solar, DGs and battery. The market is simulated over a $T=$ 24 hour scheduling horizon with a time resolution of $\Delta t=1$ hour. 
%The initial ADMM penalty parameter ($\rho$) 2.2, 
The prescribed tolerance values $(\varepsilon^{pri}, \varepsilon^{dual})$ is $10^{-3}$. The linear cost coefficient of each DG is selected within the range of $2$--$6$ Rs/kWh, while the quadratic cost coefficient is chosen within the range of $(3$--$9)\times10^{-4}$ Rs/kWh$^{2}$ \cite{wood2013power}. Table \ref{tab:system_parameters} shows the other MG specifications. For attacker, the temporal stealth between two consecutive attacks ($\Delta_i^{att}$) is 5 kW.
% the value of weighting parameters used for balancing economic impact and attack stealth are 0.9 and 0.3 respectively. 

\begin{table}[h!]
\centering
\caption{MicroGrid Specifications}
\label{tab:system_parameters}
\renewcommand{\arraystretch}{1.25}
\begin{tabular}{p{5cm} |p{2.8cm}}
\hline
\hline
Load demand & 20--60 kW \\
PV capacity & 15--30 kW \\
Maximum power trade in P2P market & 55 kW \\
DG capacity & 60--80 kW \\
Charging/Discharging power of BESS & 6--10 kW \\
Minimum SoC of BESS & 8--12 kWh \\
Maximum SoC of BESS & 33--52 kWh \\
BESS Charging/Discharging efficiency  & 95\% \\
Battery degradation cost  & $2\times10^{-6}$ Rs/kWh$^2$\\
\hline 
\hline
\end{tabular} 
\end{table}
\subsection{Dataset Generation for Detection}
\begin{figure}[h!]
\vspace{-0.4cm}
    \centering
    \includegraphics[width=\linewidth]{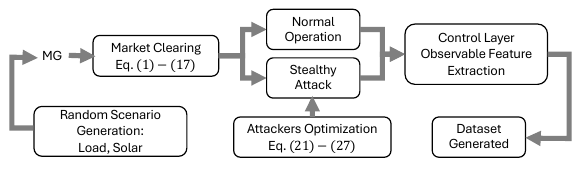}
    \caption{Dataset Generation}
    \label{fig:dataset}
\end{figure}

The dataset available to the market layer is constructed from diverse operating scenarios with varying system conditions, as illustrated in Fig. \ref{fig:dataset}. The load demand $(l_{i,t})$ is varied $\pm 30\%$, PV generation $(r_{i,t})$ is varied $\pm 20\%$, no of compromised MGs and attack budget are varied across scenarios, while the network topology and the number of participating MGs are held constant. For each scenario, ADMM-based P2P market-clearing process (Algorithm \ref{alg:admm}) is executed under both normal and attacked conditions. During each execution, the market-layer observable variables defined in Eq.~\eqref{inpdetect}, namely the primal residual $(r_i)$, dual residual $(s_i)$, graph-theoretic metrics $(\phi(G))$ and dual variables $(u_i)$, are recorded. Finally, these features are assigned the corresponding normal or attack labels to construct the supervised training dataset.

A dataset of 70,000 operating scenarios is generated, out of 35,000 samples ($N$) are normal and rest 35,000 ($A$) are attacked samples. The split of 70 \% and 30 \% are done for training and testing, respectively. 
\vspace{-0.2cm}
\subsection{Detector Specifications}
\begin{figure}[b]
\vspace{-0.2cm}
    \centering
    \includegraphics[width=\linewidth]{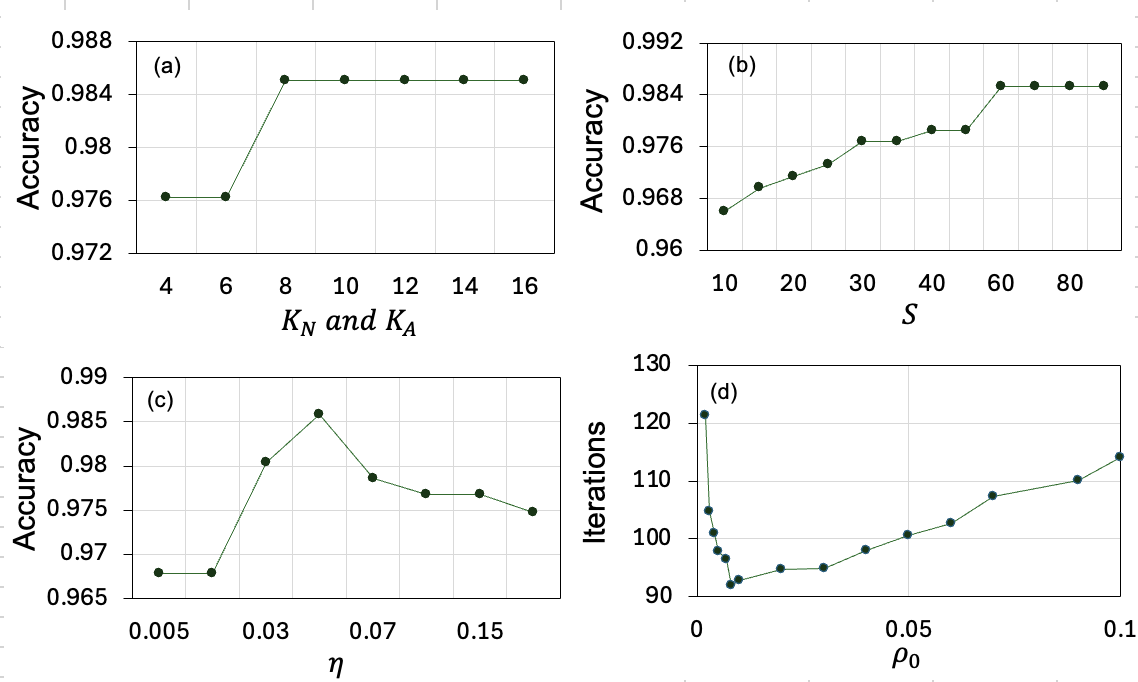}
    \caption{Sensitivity Analysis}
    \label{fig:sensitivity}
\end{figure}
The normal and attack training samples are independently partitioned into $K_N=8$ and $K_A=8$ clusters using the K-means, as increasing the number of clusters beyond eight yields negligible improvement in detection accuracy as shown in Fig. \ref{fig:sensitivity}(a). The resulting cluster centers serve as representative reference for Mahalanobis distance computation. The Gradient Boosting classifier  consists of 60 boosting stages, each employing a regression tree, where the number of stages is selected through sensitivity analysis since increasing the number of trees beyond 60 yields negligible improvement in classification accuracy (Fig. \ref{fig:sensitivity}(b)).The learning rate is chosen as 0.05, as it achieves the highest classification accuracy (Fig. \ref{fig:sensitivity}(c)).
%The Gradient Boosting classifier consists of 60 boosting stages, where each stage employs a regression tree with a maximum depth of 2 and a minimum leaf size of 20 samples.
In the mitigation process, the nominal penalty parameter $\rho_0$ is selected based on the sensitivity analysis shown in Fig.~\ref{fig:sensitivity}(d), corresponding to the minimum ADMM convergence iterations. The $\pi$ is taken as 0.5 and $\epsilon$ is taken as $10^{-8}$ for numerical stability. The smoothing factor $\lambda_\rho=0.6$ was adopted as a parameter to balance the rapid adaptation of the penalty parameter by preventing abrupt variations in rho. 
 
\section{Results and Discussion}
\label{section6}
\begin{figure}[h!]
\vspace{-0.4cm}
    \centering
    \includegraphics[width=\linewidth]{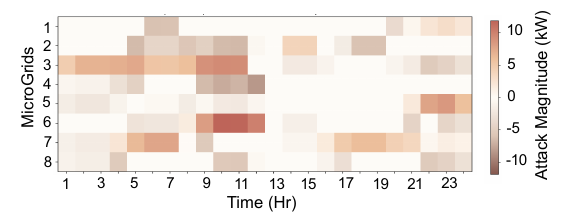}
    \caption{Attack magnitude injected by the adversary in the P2P energy trading market}
    \label{fig:cvector}
\end{figure}

This section evaluates the performance of the proposed frameworks through comprehensive simulation studies. The results are organized into three scenarios to evaluate: (i) the impact of the proposed optimal stealthy attack on the P2P energy trading market, (ii) the effectiveness of the proposed attack detection algorithm in identifying stealthy attacks, and (iii) the ability of the detector-guided adaptive-rho mitigation strategy to restore secure and reliable market operation. Fig. \ref{fig:cvector} illustrates the optimal attack magnitude injected into the P2P trading decisions of the eight MGs over the 24-hour scheduling horizon, which is the result of the optimization solved by the attacker (\ref{SECTIONIV}). 
\begin{figure}[b]
\vspace{-0.3cm}
    \centering
    \includegraphics[width=\linewidth]{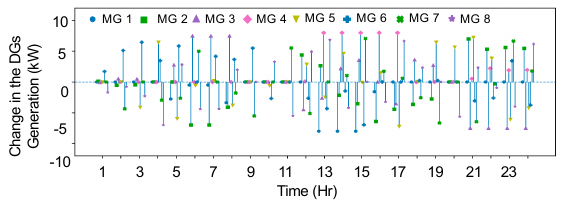}
    \caption{Change in the DGs Generation of 8 MGs due to an optimal stealthy attack}
    \label{fig:DG}
\end{figure}
\begin{figure}[b]
    \centering
    \includegraphics[width=\linewidth]{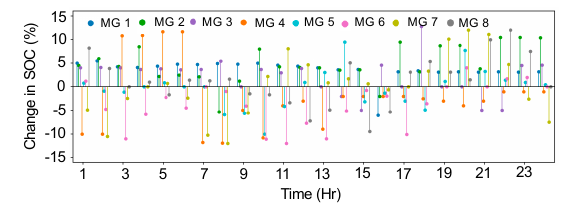}
    \caption{Change in the State of Charge (\%) of 8 MGs due to an optimal stealthy attack}
    \label{fig:SOC}
    \vspace{-0.3cm}
\end{figure}
\subsection{Scenario 1: P2P Market Operation Under Optimal Stealthy Attack}
In this section, we evaluate the effectiveness of the optimal stealthy attack. It is analyzed in terms of (i) its impact on the MG performance, (ii) its impact on the trading network structure, and (iii) its ability to maintain the stealth characteristics.

\begin{figure}[t]
\vspace{-0.2cm}
    \centering
    \includegraphics[width=\linewidth]{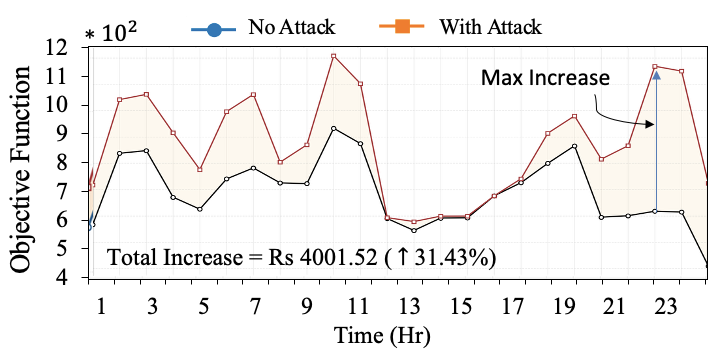}
    \caption{Change in the payoff function for 24 hours of time period due to an optimal stealthy attack}
    \label{fig:objective}
\end{figure}
\subsubsection{Impact of the MG Performance}
The modifications in the DG generation shown in Fig. \ref{fig:DG} are accompanied by corresponding changes in the battery charging and discharging schedules, as reflected by the SoC variations in Fig. \ref{fig:SOC}. Together, these results demonstrate that the stealthy attack influences both the instantaneous generation dispatch and the temporal energy management strategy of the participating MGs. 

The proposed attack also has a significant impact on the economic performance of the P2P market. The payoff function exhibits an increase of Rs. 4001.52, corresponding to approximately 31.43\% under attack conditions, indicating a degradation in system performance. Fig. \ref{fig:objective} shows the variation of objective function throughout a duration of 24 hrs. The attack exhibits its most severe impact during Hour 22, where the market objective increases by 550.26 units compared to the normal operating condition, indicating that the attack causes the maximum economic degradation during this period.
\begin{figure}[h!]
    \centering
    \includegraphics[width=0.8\linewidth]{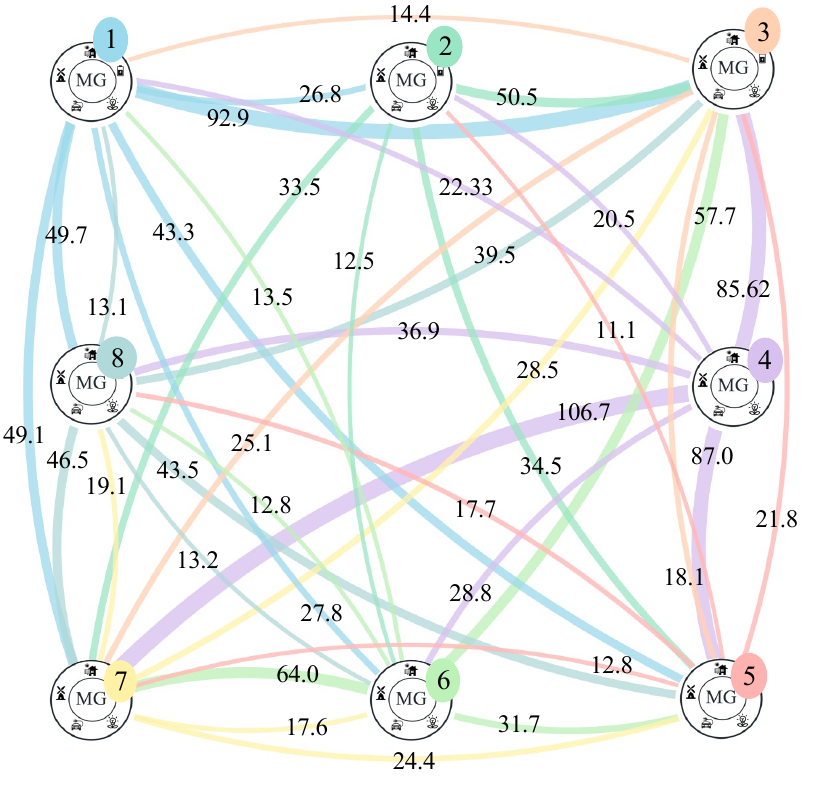}
    \caption{A trading graph for P2P energy trading among MGs}
    \label{fig:tradinggraph}
\end{figure}

\subsubsection{Impact on the Trading Network Structure}

Fig. \ref{fig:tradinggraph} illustrates the weighted directed P2P trading graph under normal operating conditions. This graph provides a structural representation of the P2P electricity market by capturing both the topology of the trading relationships and the magnitude of energy exchanged among market participants. Each node represents a participating MG, while each directed edge denotes a bilateral energy transaction between the two MGs. The numerical value associated with each edge indicates the cumulative energy traded over the scheduling horizon, and the edge thickness is proportional to the corresponding transaction magnitude.
\begin{table}[h!]
\centering
\caption{Comparison of Graph Metrics Under Normal and Optimal Attack Conditions}
\label{tab:graph_metrics}
\renewcommand{\arraystretch}{1.25}
\begin{tabular}{l c c c}
\hline    
\hline    
Graph Metric &  Expression& No Attack & Attack \\
\midrule

Total P2P Trade &
$\displaystyle \sum_{i=1}^{M}\sum_{j=1}^{M}F_{ij}$ &
2270.88 & 2679.61 \\

Directed Edges &
$\displaystyle \sum_{i=1}^{M}\sum_{j=1}^{M}\mathbf{1}(F_{ij}>0)$ &
50 & 52 \\

Max Out-Degree &
$\displaystyle \max_{i}\left(\sum_{j=1}^{M}F_{ij}\right)$ &
910.24 & 789.28 \\

Max In-Degree &
$\displaystyle \max_{j}\left(\sum_{i=1}^{M}F_{ij}\right)$ &
620.85 & 731.58 \\

Frobenius Norm &
$
\sqrt{\sum_{i=1}^{M}\sum_{j=1}^{M}F_{ij}^{2}}$ &
500.45 & 525.24 \\

Spectral Radius &
$\displaystyle \max_i |\lambda_i(F)|$ &
153.02 & 204.78 \\
\hline
\hline
\end{tabular}
\end{table}

Consequently, the weighted adjacency matrix, as shown in Table \ref{tab:graph_metrics}, captures the structural characteristics of the P2P market and forms the basis for computing the graph-theoretic metrics used in this work, including total P2P trade, active directed edges, maximum in/out-degree, Frobenius norm, and spectral radius. The table shows how these trading matrices deviate from the attack. These results confirm that the proposed attack significantly changes the structural characteristics of the P2P trading graph.
\subsubsection{Stealthiness Analysis}
\begin{figure}[h!]
    \centering
    \includegraphics[width=\linewidth]{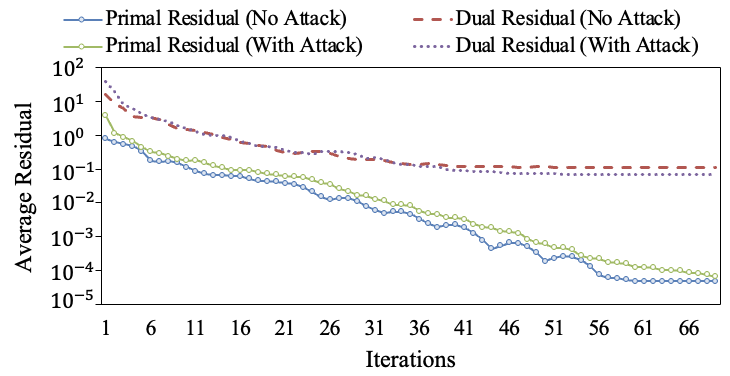}
    \caption{Primal and dual residual convergence before and after attack }
    \label{fig:attack003}
    \vspace{-0.5cm}
\end{figure}

As the attack deviates the P2P power traded among MGs, but the attacker makes sure that the stealth characteristics does not change, so that it will become difficult for the control layer to detect the attack. As shown in Fig. \ref{fig:attack003}, the primal and dual residuals continue to satisfy the convergence criteria despite the presence of the attack. Consequently, the distributed market-clearing process appears to operate normally, highlighting the stealthy nature of the proposed attack.
\subsection{Scenario 2: Detecting the Optimal Stealthy Attack}
To evaluate the effectiveness of the proposed detector in identifying the optimal stealthy false data injection attack, the following confusion matrix summarizes the classification outcomes for both normal and attack scenarios.

\begin{equation}
\mathbf{CM}=
\begin{bmatrix}
10399 & 101 \\
216 & 10284
\end{bmatrix}
\label{eq:confusion_matrix}
\end{equation}

where the first row corresponds to normal operating conditions and the second row corresponds to attack scenarios. The detector correctly identifies 10,399 normal scenarios (True Negatives $TN$) while incorrectly classifying only 101 normal scenarios as attacks (False Positives $FP$). Similarly, 10,284 attack scenarios are correctly detected (True Positives $TP$), whereas only 216 attack scenarios remain undetected (False Negatives $FN$). These results indicate that the proposed detector achieves a very low false alarm rate while simultaneously maintaining a high attack detection capability.

Table \ref{tab:detection_comparison} compares the classification performance of the proposed Mahalanobis--Boosting detector with RBF-SVM, conventional Gradient Boosting (GB), SVM, and XGBoost. All classifiers are trained and evaluated using the same input feature set (Eq.\ref{inpdetect}), while the proposed method additionally constructs Mahalanobis representations from these features to explicitly capture covariance-aware inter-feature dependencies. Here, accuracy is computed as $(TP+TN)/(TP+TN+FP+FN)$, precision as $TP/(TP+FP)$, recall as $TP/(TP+FN)$, and the F1-score as $2TP/(2TP+FP+FN)$. The proposed method achieves the highest accuracy, precision, recall, and F1-score, outperforming the benchmark methods that do not effectively model covariance-aware inter-feature dependencies. The resulting performance improvement therefore 
demonstrates the contribution of explicitly incorporating covariance-aware class geometry into the boosting-based detection framework.

\begin{table}[h!]
\centering
\caption{Performance Comparison of Different Detection Methods}
\label{tab:detection_comparison}
\setlength{\tabcolsep}{4pt}
\renewcommand{\arraystretch}{1.2}
\begin{tabular}{
>{\RaggedRight\arraybackslash}p{2.2cm}|
>{\centering\arraybackslash}p{1.2cm}
>{\centering\arraybackslash}p{1.2cm}
>{\centering\arraybackslash}p{1.2cm}
>{\centering\arraybackslash}p{1.2cm}}
\hline
\hline 
Method &
Accuracy &
Precision &
Recall &
F1-Score \\
\hline
Proposed Method
& 98.49\%
& 99.03\%
& 97.94\%
& 98.48\% \\
RBF-SVM
& 90.56\%
& 91.98\%
& 88.43\%
& 90.12\%\\
Gradient Boosting
& 85.01\%
& 88.21\%
& 80.82\%
& 84.32\% \\
SVM
& 86.26\%
& 88.41\%
& 83.27\%
& 85.78\% \\
XGBoost
& 85.53\%
& 89.15\%
& 81.70\%
& 85.32\% \\
\hline
\hline
\end{tabular}
\end{table}

\begin{table*}[b!]
\centering
\caption{Comparison of Convergence and Residual Characteristics Under Different Cases}
\label{tab:convergence}

\renewcommand{\arraystretch}{1.5}
\setlength{\tabcolsep}{10pt}

\begin{tabular*}{\textwidth}{@{\extracolsep{\fill}}lccccc}
\hline
\hline
Scenario
& Operating Cost (Rs)
& Iterations 
& Convergence Time (seconds)
& Primal Residual 
& Dual Residual\\
\hline

Case 1: Normal 
& 12728.13
& 970
& 0.5837 
& $4.87409 \times 10^{-5}$ 
& 0.043218\\

Case 2: With Attack
& 16729.66 
& 1121
& 0.8726 
& $6.63524 \times 10^{-5}$ 
& 0.069087\\
Case 3: With Mitigation (Proposed)
& 13644.55 
& 1029
& 0.6496 
& $5.19457 \times 10^{-5}$ 
& 0.065757\\
\hline
\hline
\end{tabular*}
\end{table*}

\begin{figure}[t]
    \centering
    \includegraphics[width=0.8\linewidth]{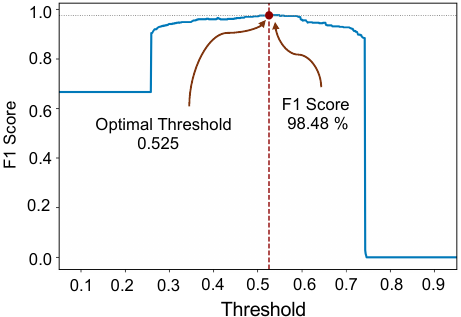}
    \caption{Selection of the optimal threshold using best F1 score}
    \label{fig:threshold}
    \vspace{-0.2cm}
\end{figure}

Fig. \ref{fig:threshold} illustrates the variation of the F1-score with the decision threshold used for attack classification. The threshold is varied over the interval [0,1], and the corresponding F1-score is evaluated on the validation dataset. The maximum F1-score of 98.48\% is achieved at a threshold of 0.525, which is therefore selected as the optimal decision threshold. This threshold provides the best balance between precision and recall, ensuring high attack detection performance while minimizing both false positives and missed detections.

\begin{figure}[h!]
\vspace{-0.2cm}
    \centering
    \includegraphics[width=\linewidth]{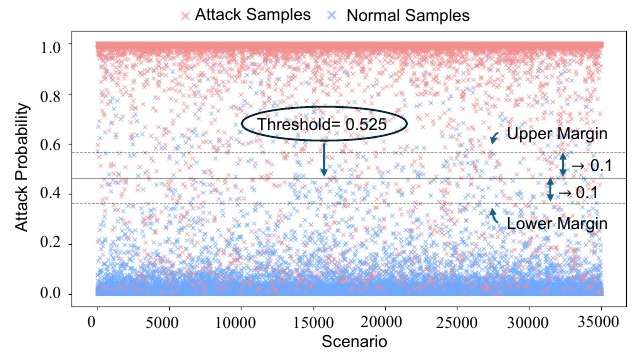}
    \caption{Attack Probability}
    \label{fig:prob}
    \vspace{-0.2cm}
\end{figure}

Based on the optimal threshold calculated, Figure \ref{fig:prob} illustrates the predicted attack probabilities for all evaluated scenarios. Each point corresponds to one operating scenario, where blue markers represent normal samples and red markers represent attack samples. The solid horizontal line denotes the optimal decision threshold, while the dashed lines indicate the upper and lower suspicious margins used to identify uncertain operating regions.
It can be observed that the majority of normal scenarios are assigned probabilities close to zero, whereas attack scenarios are concentrated near one, indicating a clear separation between the two classes. Only a small number of samples appear within the suspicious margin surrounding the decision threshold, suggesting that the learned decision boundary effectively discriminates between normal and malicious operating conditions. The distinct clustering of the two classes explains the high classification accuracy achieved by the proposed detector.
\subsection{Scenario 3: Detector-Guided Adaptive-rho Mitigation}
Table \ref{tab:convergence} compares the convergence characteristics under normal operation, stealthy attack, and the proposed detector-guided adaptive-rho mitigation strategy. The stealthy attack increases the operating cost from Rs 12,728.13 to Rs 16,729.66 (31.4\%), while increasing the cumulative number of ADMM iterations over the complete evaluation horizon and convergence time to 1121 and 0.8726 s, respectively. After applying the proposed mitigation strategy, the operating cost is reduced to Rs 13,644.55, with the number of iterations and convergence time decreasing to 1029 and 0.6496 s, respectively. These results demonstrate that the proposed approach effectively mitigates the impact of the attack while restoring the convergence performance close to normal operating conditions.
\section{Conclusion}
\label{section7}

This paper presented an Mahalanobis–Boosting detector for identifying stealthy false data injection attacks in distributed P2P energy trading. By iteratively refining the Mahalanobis feature representation through hardness-weighted prototype and covariance updates, the proposed framework progressively enhances the discrimination between normal and attack samples. The detector achieved an accuracy of 98.49\%. Furthermore, the estimated attack probability was integrated into an attack-aware adaptive ADMM framework, which saves the operating cost by around 18.43\% relatuve to the attacked case, enabling secure and resilient distributed market clearing under cyber-attacks. The proposed framework provides an effective and privacy-preserving solution for enhancing the cyber resilience of distributed P2P energy trading markets.

%Future work will investigate the extension of the proposed framework to network-constrained P2P energy trading, adaptive adversaries with evolving attack strategies, and distributed multi-cloud architectures for large-scale distribution systems.

\begin{small}
\bibliographystyle{IEEEtran}
\bibliography{references}
\vfill
\end{small}
\end{document}